**Materials Swelling Revealed Through Automated Semantic Segmentation of Cavities in Electron Microscopy Images**

Ryan Jacobs[1], Priyam Patki[2], Matthew Lynch[2], Steven Chen[2], Dane Morgan[1], Kevin G. Field[2]

[1]Department of Materials Science and Engineering, University of Wisconsin-Madison, Madison, Wisconsin, 53706, USA

[2]Nuclear Engineering and Radiological Sciences, University of Michigan - Ann Arbor, Michigan, 48109 USA

[†]Corresponding author email: rjacobs3@wisc.edu

# Abstract

Accurately quantifying swelling of alloys that have undergone irradiation is essential for understanding alloy performance in a nuclear reactor and critical for the safe and reliable operation of reactor facilities. However, typical practice is for radiation-induced defects in electron microscopy images of alloys to be manually quantified by domain-expert researchers. Here, we employ an end-to-end deep learning approach using the Mask Regional Convolutional Neural Network (Mask R-CNN) model to detect and quantify nanoscale cavities in irradiated alloys. We have assembled the largest database of labeled cavity images to date, which includes 400 images, >34k discrete cavities, and numerous alloy compositions and irradiation conditions. We have evaluated both statistical (precision, recall, and F1 scores) and materials property-centric (cavity size, density, and swelling) metrics of model performance, and performed in-depth analysis of materials swelling assessments. We find our model gives assessments of material swelling with an average (standard deviation) swelling mean absolute error based on random leave-out cross-validation of 0.30 (0.03) percent swelling. This result demonstrates our approach can accurately provide swelling metrics on a per-image and per-condition basis, which can provide helpful insight into material design (e.g., alloy refinement) and impact of service conditions (e.g., temperature, irradiation dose) on swelling. Finally, we find there are cases of test images with poor statistical metrics, but small errors in swelling, pointing to the need for moving beyond traditional classification-based metrics to evaluate object detection models in the context of materials domain applications.

## 1     Introduction

Metal alloys used in nuclear reactor cores and surrounding structures undergo irradiation, causing damage to the material which can result in the production of extended defects such as dislocation loops, precipitates, and cavities (sometimes called voids when they do not contain gas or bubbles when they do contain gas) that, in turn, have a deleterious impact on the mechanical properties via hardening, embrittlement and swelling.[1–5] Bias-driven growth of cavities leading to unconstrained swelling under neutron irradiation generally occurs via the presence of helium (produced from nuclear transmutation) that stabilizes the cavities.[3,6] Significant swelling can result in material degradation and failure, hence, understanding the interplay of alloy composition, microstructure, and reactor conditions such as operating temperature and irradiation dose are important for informing safe and reliable reactor operation.[7] Bulk measurement methods of reactor components, such as the Archimedes method, are typically easiest to conduct to obtain information on the total volumetric swelling response of a material.[8] However, Transmission and Scanning Transmission Electron Microscopy (S/TEM) methods are also commonly employed in materials research and development evaluations for *ex situ* characterization of alloy microstructure and swelling quantification. TEM methods have an advantage over bulk measurement methods as they enable one to obtain the strict swelling response from the presence of cavities, eliminating swelling contributions from other factors such as creep, secondary phase formation, and phase densification at high temperature. TEM analysis can also be used to identify swelling responses locally, e.g., as is seen during ion irradiations or in complex microstructures due to localized microstructural effects on the helium and defect formation energetics and kinetics. Finally, TEM analysis can be used to help understand early stage irradiation response, e.g., the nucleation and growth process of cavities, which initiates before significant macroscopic swelling has occurred. Such microscale characterization thus enables detailed mechanistic understanding important for the design of swelling resistant alloys, and enables researchers to understand linkages between material microstructure, composition, and swelling response as a function of key operational variables such as temperature, irradiation type (e.g., neutron vs. ion), dose rate, and total dose.[9] This information is in turn useful for informing materials modeling of swelling in different regimes (i.e., incubation, transient, and steady state swelling) and can help inform operational limits of a material in a nuclear reactor.[5]

At present, swelling quantification from TEM samples is typically performed by considering a handful of TEM images and manually counting and measuring individual cavities in each image, for example using image analysis programs such as ImageJ.[10] This approach typically treats relatively small sample sizes due to (1) the time and resource-intensive nature of TEM sample preparation and (2) the cavity labeling and counting analysis. Regarding the first issue, recent advances in TEM sample preparation, including high-throughput focused ion beam (FIB) methods (e.g., plasma FIB) and flash polishing, can be used to generate an extensive library of TEM samples.[11,12] Therefore, sample preparation limitations are rapidly being overcome. We note also that modern TEM instruments have undergone exponential growth in data acquisition rates with the development of new detector technologies, resulting in higher resolution images and larger overall data sizes.[13–16] Therefore, it is clear that manual labeling and measurement of cavities will not be able to keep pace with the scaling of TEM dataset sizes. Thus, the second issue above is rapidly becoming the bottleneck in scaling up image-based analysis capabilities. An automated method that can quickly analyze large TEM datasets, automatically detect and quantify cavities, and then assess material swelling would enable researchers to evaluate many more areas of interest on a given sample, providing more robust statistics, quantification of effects of heterogeneity, and in-depth evaluations of cavity properties and material swelling.

In the past decade, deep learning methods have witnessed significant advancement. They have resulted in revolutionary changes to the field of computer vision. Specifically, in the context of object detection, deep convolutional neural networks (CNNs) such as ResNet50, ResNet101 and VGG16 are used to extract detailed underlying feature sets from tens of thousands of images in canonical databases such as ImageNet[17] and Common Objects in Context (CoCo).[18] These so-called "backbone" networks are implemented in CNN-based object detection frameworks such as the Faster Regional Convolutional Neural Network[19] (R-CNN) and Mask R-CNN models,[20] which contain additional neural networks that suggest regions of interest in the image and classify and segment individual objects within each region of interest.[21,22] There has been a growing body of work applying object detection methods to electron microscopy images in materials science,[23] with applications ranging from detecting various defects (e.g., dislocations, precipitates, black dot defects) in irradiated metal alloys[24–26] to quantifying micro and nanoparticles[27,28] and finding individual atoms in high-resolution STEM images.[29,30] Most relevant to the present work, Anderson et al. used the Faster R-CNN model to detect cavities in Ni-based X-750 alloys.[31] Their Faster R-CNN model effectively found cavities, with reported F1 scores in the range of 0.7-0.8. Because the Faster R-CNN model does not provide pixel-level segmentation information, additional post-processing methods separate from the deep learning model were used to extract the cavity size information from the predicted bounding boxes. The present work employs the Mask R-CNN model to realize a fully end-to-end deep learning cavity detector. We include the publicly available data used in the work of Anderson et al. from the Canadian Nuclear Laboratory (CNL), which we refer to as the CNL dataset in this work, and significantly expand the previously available cavity image database to include images comprising a greater range of alloy compositions and irradiation conditions by including new images from the Nuclear Oriented Materials & Examination (NOME) Laboratory at the University of Michigan, which we refer to as the NOME dataset in this work (see **Section 4** for more information). Two examples of images from each of the CNL and NOME datasets are shown in **Figure 1**.

There are many possible ways of assessing a segmentation machine learning model for defects. One level is how the model performs as a classification algorithm, which can be done for any object classified by the model. A typical model provides classification for pixels (in or out of the defect), defects (found or not found), and defect types (for cases with multiple defect types). Such classification performance is generally characterized by metrics such as precision, recall, accuracy, and F1 scores. A second level of assessment is how the model performs for defect properties, which might include basic properties (e.g., size distribution, mean size, density, shape, position, etc.) and evolutions or correlations associated with those basic properties (e.g., growth rate, diffusivity, pair distribution function, etc.). A third level of assessment is materials properies, which for irradiated alloys are generally swelling or hardening predictions based on physical models and properties of the observed defects. Assessments like those just listed can generally be done with different groupings of the data, e.g., for a fixed area, on a per image basis, or for a specific set of images. Also, since assessments are generally done on left-out test data, those test data sets can be generated by different methods, the most common being choosing them at random (e.g., k-fold cross-validation) or removing specific groups of data with select properties to represent likely use cases for the model. In this work, we focus assessment on classification scores for finding defects, defect size distribution and density, and material swelling. We do this on both a per-image basis and averaged over multiple images. Together, these assessments explore the accuracy of the model for the information typically utilized by the radiation effects community.

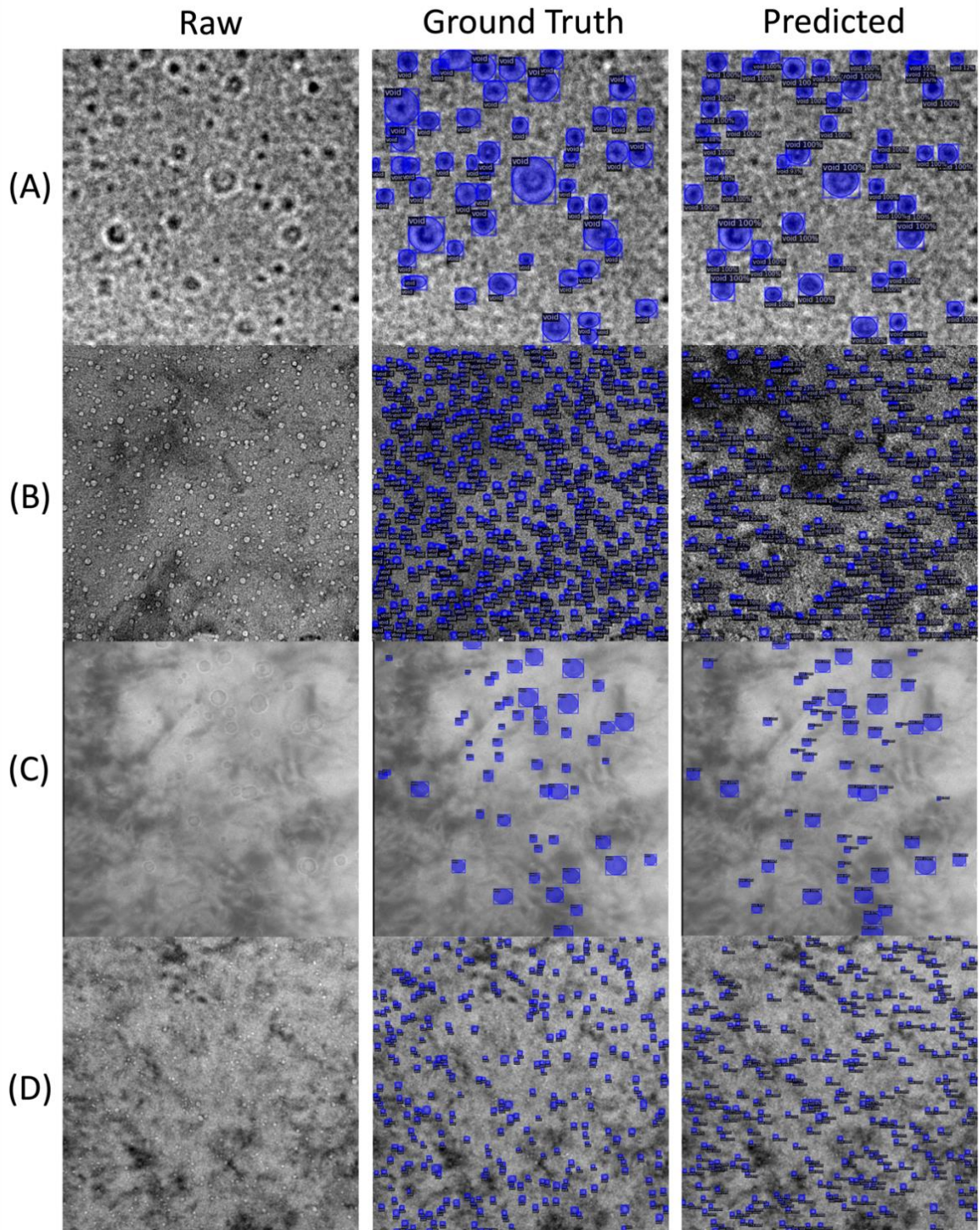

**Figure 1:** Example raw images (left column) with ground truth cavities labeled (middle column) and corresponding Mask R-CNN model predictions (right column). (A) CNL overfocused image with F1=0.82. (B) CNL underfocused image with F1=0.63. (C) NOME overfocused image with F1=0.81. (D) NOME underfocused image with F1=0.83.

## 2  Results and Discussion

### 2.1  Benchmarking model performance of assessing material swelling

Throughout this work, we focus our model assessment on its ability to assess material swelling and investigate the primary sources of error in material swelling. Here, we first benchmark the performance of Mask R-CNN models trained and tested on different random subsets of our complete CNL+NOME cavity database (see **Section 4**). Evaluation with random cross-validation forms a baseline for how well the model is expected to perform on test images that, at least qualitatively, are drawn from the same domain as the training set. **Figure 2** contains a parity plot comparing model predicted vs. true values of average per-image material swelling for five different train/test splits of the CNL+NOME database. Key fit statistics of coefficient of determination ($R^2$), mean absolute error (MAE), mean absolute percentage error (MAPE), root mean squared error (RMSE), and RMSE divided by the true dataset standard deviation (reduced RMSE, RMSE/$\sigma$) are included. A summary of the key classification metrics and materials property metrics for each split, together with the average and standard deviation across all five splits, can be found in **SI Note 1**. Regarding material swelling in **Figure 2**, across the five splits examined, the average MAE is 0.30 percent swelling with a standard deviation of 0.03 percent swelling. The best split was the CNL+NOME initial split with an MAE = 0.26 percent swelling, while the worst split was CV split 1 with an MAE = 0.35 percent swelling. In addition to assessments of material swelling, we have also provided a detailed examination of model assessments of per-image average cavity size and and cavity areal density, which is provided in **Figure S1** of **SI Note 1**. The model can assess the average per-image cavity size with high accuracy, with an average (standard deviation) MAE of just 1.02 (0.14) nm, which corresponds to an average (standard deviation) MAPE value of 8.94% (0.84%) error in cavity size, which is a similar error level as our previous work.[25] Our model has the highest errors in assessing cavity density, particularly for images with high cavity densities (> 20 ×10$^4$ nm$^{-2}$), where the model has a clear bias to lower values. The interplay of cavity size and density with regard to swelling assessments is discussed in **Section 2.3**. Overall, the Mask R-CNN model can assess the material swelling well with a typical mean absolute error of about 0.30 percent swelling, which is a small enough error for the model to discern changes in swelling repsonses based on material design (e.g., alloy refinement) and service conditions (e.g., temperature, dpa) and thus readily provides an accelerated means to assess these factors in TEM-based swelling quantification workflows.

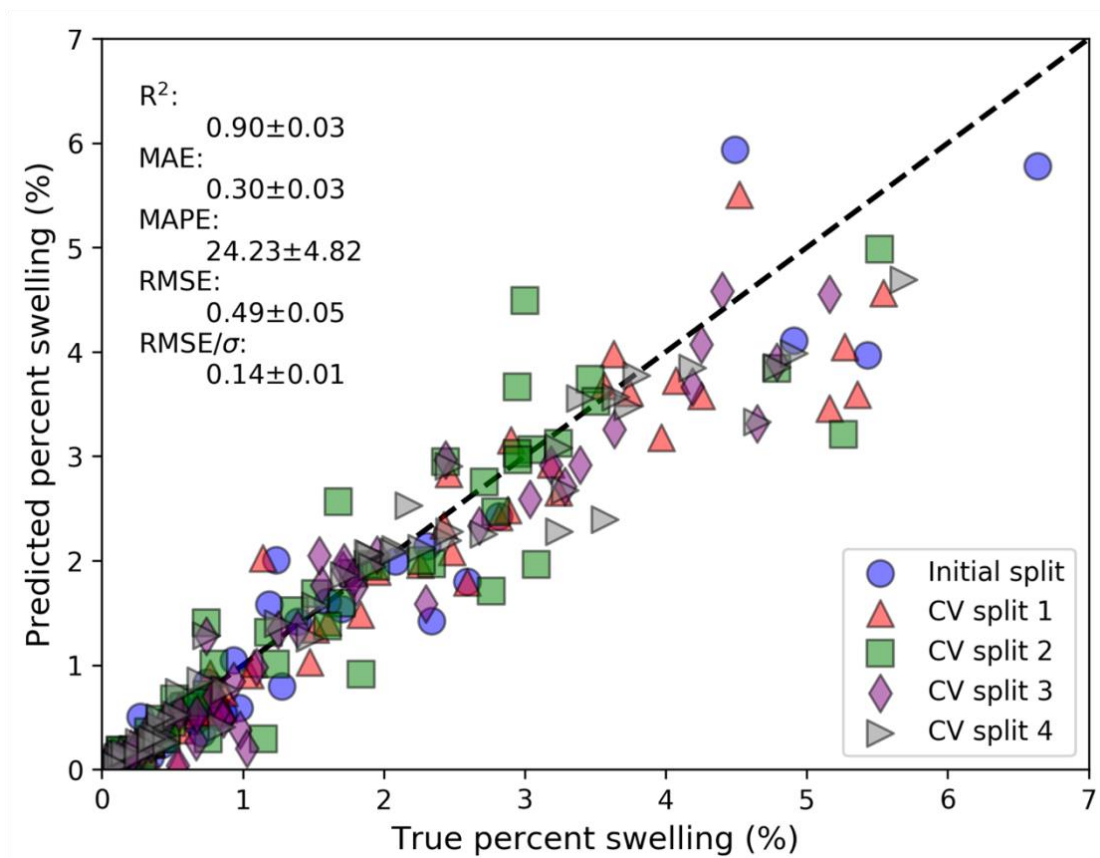

**Figure 2:** Parity plot of true and predicted material swelling. The different symbols correspond to different cross validation train/test splits. The fit statistics in black text denote the average +/- standard deviation across all five splits for each metric.

## 2.2 Model domain assessment with leave-out group tests

From the above discussion, the model trained on our complete CNL+NOME dataset yielded accurate assessments of material swelling for randomly left-out test images, constituting a test of model performance on images drawn qualitatively from the same domain as the training set. A more demanding test of the ability of our Mask R-CNN models to assess material swelling involves testing the model on images quite distinct from those encountered in training by the use of leave-out group cross validation. While there are many ways one can leave out physically motivated groups of data, here we focus on the practical scenario of applying our trained Mask R-CNN to cavity images belonging to a distinct dataset from that used in training. To do this, we train a model solely on the CNL data, use it to predict CNL and NOME test data, and compare it to the model trained on the combined CNL+NOME dataset from **Section 2.1**. Likewise, we also train a model only on the NOME data, and use it to predict CNL and NOME test data.

**Figure 3** contains parity plots of material swelling assessment for our leave out group cross validation test. A detailed summary of the materials property statistics (cavity size, density, and swelling values) for the tests shown in **Figure 3** can be found in **SI Note 2**. In **Figure 3A,** the model is trained on only CNL data and is used to predict swelling on CNL test images (blue points) and NOME test images (red points). The CNL and NOME test points are separated based on whether the test images correspond to overfocused (circle symbols) or underfocused (triangle symbols) imaging conditions where the different conditions invert the contrast modulation of the cavities present in the material. In **Figure 3A**, we see that the model trained on CNL images demonstrates good assessment of material swelling on the CNL image test set with an MAE of 0.40 percent swelling. The model performs better on underfocused images compared to overfocused images from the standpoint of MAE, where the swelling MAE values on underfocused (overfocused) images are 0.33 (0.53) percent swelling, respectively (see **SI Note 2**). The improved performance on underfocused images is likely due to their being more underfocused versus overfocused cavities in the CNL database. A similar response was observed in our previous work using Mask R-CNN to detect dislocation loops in FeCrAl alloys, where our learning curves showed best model performance on the defect types present in highest quantity in the training data.[25] In **Figure 3A**, we can also see that the model trained on CNL data performs poorly on the NOME test set. While at first glance the MAE value of 0.66 percent swelling on the NOME test set does not appear much worse than the MAE of 0.40 percent swelling on the CNL test set, the range of swelling values for the NOME data are much smaller, and the higher error, in this case, is better exemplified by inspecting the MAPE value of about 215% for testing on NOME vs. just under 20% for testing on CNL, as well as the reduced RMSE value which is much higher (lower) than unity for the NOME (CNL) test set.

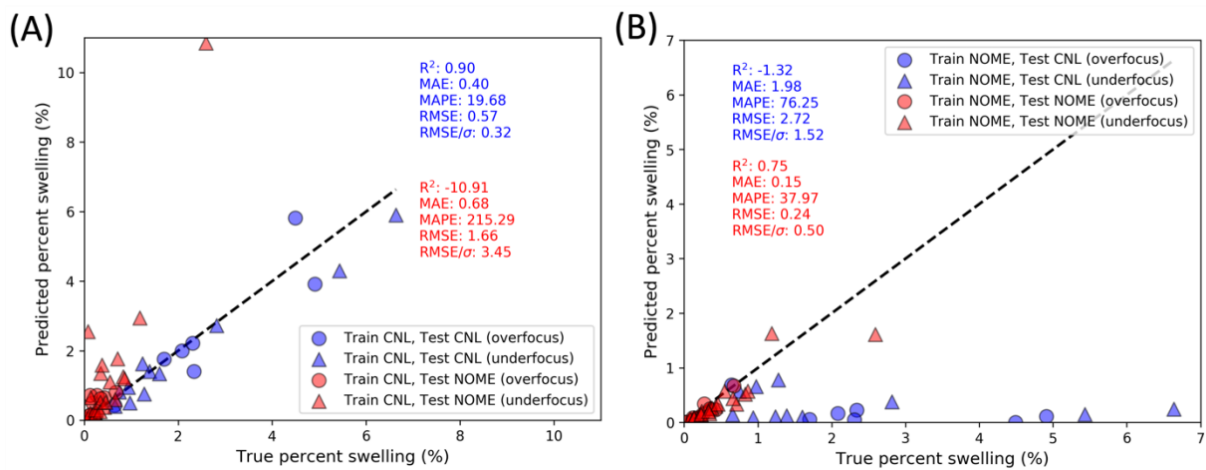

**Figure 3**: Parity plots assessing Mask R-CNN per-image performance of predicting materials swelling. (A) CNL initial split, with model trained on CNL and tested on CNL (blue data) and trained on CNL and tested on NOME (red data). (B) NOME initial split, with model trained on NOME and tested on CNL (blue data), and trained on NOME and tested on NOME (red data). In both plots, the circle and triangle points denote overfocused and underfocused images, respectively, and the color-coded fit statistics coincide with the corresponding set of points of like color.

In **Figure 3B**, we perform the test case where the model is trained only on the NOME data and separately tested on the CNL and NOME test sets. The model trained and tested on NOME data shows an excellent overall ability to assess swelling, with an MAE of just 0.15 percent swelling (MAPE = 37.97%). In contrast, the model performs poorly on assessments of the CNL test set, with large swelling MAE (MAPE) values of 1.98 percent swelling (76.25%), respectively, and essentially no ability to assess swelling of samples with true swelling values greater than about 1.5 percent swelling. This result makes sense through the lens of model applicability domain. While the NOME dataset constitutes a more diverse set of alloy compositions and irradiation conditions, the swelling present in the NOME images has a maximum of about 2.5 percent swelling (with all but one test image having less than 1.5 percent swelling), in contrast to the large swellings of some CNL images of up to nearly 7 percent swelling.

We reiterate that by training a model which uses both the CNL and NOME data (**Figure 2** and **Figure S2 in SI Note 2**), the model provides an accurate assessment of material swelling both on the separate CNL (MAE = 0.44 percent swelling) and NOME (MAE = 0.15 percent swelling) test sets, and collectively shows an MAE of 0.26 percent swelling. The model trained on CNL and NOME data shows virtually unchanged performance on each test subset compared to individually trained models shown in **Figure 3** and **Figure S2**, indicating that the combined model has a larger applicability domain. The combined CNL+NOME model shows approximately identical performance predicting swelling of overfocused (0.26 percent swelling) vs. underfocused (0.27 percent swelling) images, though from the standpoint of MAPE the model performs better on underfocused images (29.03 %) compared to overfocused images (39.13 %) (see **Table S2** in **SI Note 2**). In addition to the materials property statistics summarized here, we have collected the classification statistics of overall P, R, and F1 scores and average per-image P, R, and F1 scores for the tests discussed above. We find that the conclusions regarding model performance in the context of material swelling generally persist when considering the overall and average per-image F1 scores. Additional discussion of the classification metrics and a table of their values can be found in **SI Note 2**. Overall, our results demonstrate that it is preferable to simply train one model with training images from both datasets, as the model domain is widened without loss in classification or materials property metric performance within any given single dataset.

## 2.3 Understanding model errors of swelling assessment

Here, we seek to better understand the source of error in the model swelling assessments. Based on the equation to calculate material swelling (Eq. 1, see **Section 4**), it is intuitive that cavity size (cubic scaling) has a larger impact than cavity density (linear scaling) to determine the swelling (see **SI Note 3** for a visualization of this fact using our present database). Given the detailed data obtained from the Mask R-CNN model output, we show this effect in practice and quantify potential problematic areas of model use more precisely. **Figure 4A** shows the relationship between the true per-image cavity size and the model *error* in the cavity density. In **Figure 4A** the sizes of the data points scale with the model *error* in the swelling. What we learn from **Figure 4A** is that the images with the highest density errors are those with small cavities, at least on average. The small sizes of the points with high density errors indicate that these images with poor density assessments also have minor swelling errors. From the standpoint of desiring a model which produces accurate swelling assessments, the fact that at times the model shows poor assessments of cavity density are not necessarily concerning, as the poor density assessments coincide with small swelling errors, at least for the images analyzed in our present database. It is worth noting that our model is largely unbiased with regard to cavity size predictions (see **Figure S1A** in **SI Note 1**), biased to underpredict cavity densities (see **Figure S1B** in **SI Note 1**), resulting in essentially no bias in the swelling errors (see **Figure 2**), which is due to the fact that small cavities have a small impact on the swelling values, and are the cavities that are undercounted in the density predictions. In **Figure 4B**, we plot the average absolute swelling error as a function of the true per-image cavity size, binned based on ranges of cavity sizes. The sizes of the points in **Figure 4B** correspond to the number of test images contained in each cavity size bin. The error bars denote the standard error in the mean of the average absolute swelling error in each cavity size bin. As an example, to obtain the first data point of the 0-5 nm binned NOME data, the sizes of the red square points in **Figure 4A** that are between 0-5 nm on the y-axis are averaged to obtain the average absolute swelling error in **Figure 4B,** the error bar is the standard error in the mean of those same points, and the size of the point in **Figure 4B** scales with the number of data points in the 0-5 nm size bin (note this is why larger points tend to have smaller error bars). In **Figure 4B**, we can see that the CNL (NOME) images with average cavity sizes greater than 10 nm (15 nm) have higher average swelling errors than the overall MAE of 0.3 percent swelling from random cross validation. Taken together, the analysis shown in **Figure 4** points to images with large cavities being the most susceptible to high swelling errors, with errors potentially twice as high as that obtained from our random cross validation test. As a further piece of analysis, in **Figure S4** in **SI Note 3** we have additional plots like that shown in **Figure 4B**, except we plot the average absolute swelling error as a function of the (binned) true swelling, for the cases of all test images together as well as split out by CNL and NOME subsets. This analysis indicates we have smaller (larger) absolute swelling errors (percentage swelling errors) when the true swelling is small (e.g., average swelling error of 0.13% and percentage error of 33.0% for true swelling <1%) and larger (smaller) absolute swelling errors (percentage swelling errors) when the true swelling is large (e.g., average swelling error of 0.60% and percentage error of 16.0% for true swelling >2%). Overall, across all test images in our database, our model shows average absolute swelling errors (percentage swelling errors) of about 0.3% (25%).

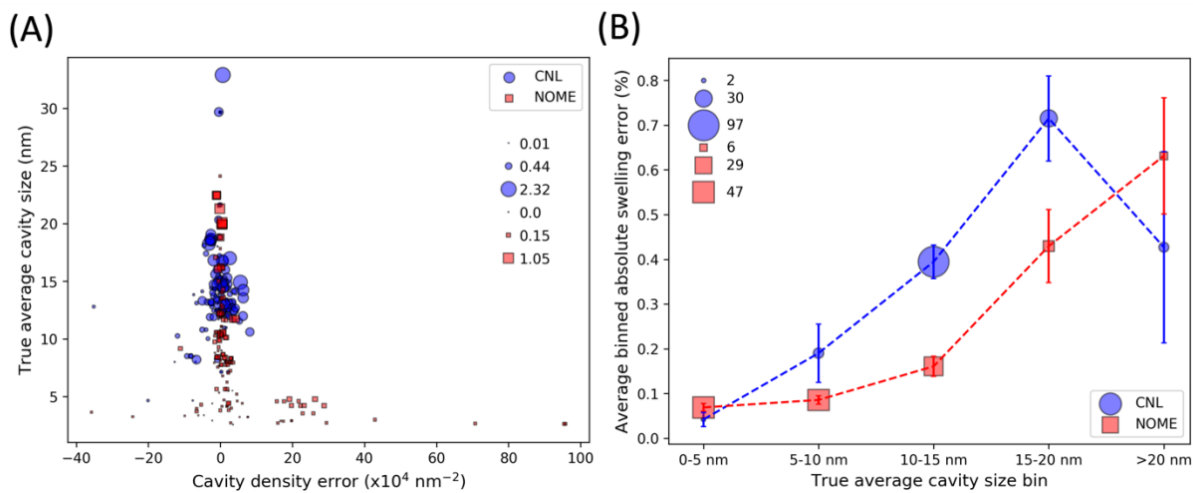

**Figure 4:** (A) Relationship between the true per-image cavity size and the model error in assessing the corresponding cavity density. Each data point represents one test image, where the blue circles and red squares denote CNL and NOME test images, respectively. The size of the data points scales with model error of the percent swelling. (B) The trend of model predicted absolute error in material swelling as a function of true average cavity size. Here, the x-axis represents binned values of true cavity size (i.e., groups of test images based on their range in true cavity sizes from the y-axis of the plot (A). The blue circles and red squares denote groups of CNL and NOME test images, respectively. The size of the points scales with the number of test images comprising the true average cavity size bin. The size legends denote the minimum, average, and maximum for the respective data trace. The error bars are the standard error in the mean of the absolute swelling error.

The above discussion focused on correlating average values of cavity size with errors in cavity density and swelling. As a final piece of analysis, we examine two specific images in greater detail to better understand the impact of the entire cavity size distribution on the swelling error. These particular cases were test images from the CNL+NOME initial split case. We examine two extreme cases for this analysis: an underfocused NOME test image named "10 59 K.png" which showed the smallest swelling error of just 0.004 percent swelling, and an underfocused CNL test image name "02.jpg" which showed the largest swelling error of 1.46 percent swelling. The ground truth and model predicted cavity labels are shown in **Figure 5**. As a first remark, the NOME image with the best swelling assessment showed a low F1 score of just 0.50, while the CNL image with the worst swelling assessment showed a high F1 score of 0.90. This finding points to the importance of evaluating materials property-centric metrics in addition to, or as a substitute for, conventional classification-based metrics for cases where the model is being evaluated for use in a specific materials domain application. Next, the cavity size distributions (represented here as fraction of the image size) shown in **Figure 5** highlight that the poor F1 of the NOME image is the result of the model missing many small cavities spanning about 2% or less of the image size (for this particular image, this amounts to cavities about 5 nm in size), while for the CNL image a handful of large cavities are missing or found cavities were predicted to have underestimated sizes. For the NOME image in **Figure 5A**, the small swelling error is the result of a slight overestimation of the average cavity size, driven mainly by the prediction of a single large cavity with a size of about 5% of the image, as shown by the cumulative swelling contributions overlaid with the size histogram. In **Figure 5B**, the poor swelling assessments are driven by the model missing the largest cavities in the image, and the over-representation of small predicted cavities does not make up for the underpredicted swelling. Overall, the extreme examples presented in **Figure 5** show that, at times, our model can show a good swelling assessment that is the result of error cancellation (**Figure 5A-** misses many small cavities but has one large false positive cavity) and our model can have a poor swelling assessment that is the result of a combination of errors (**Figure 5B-** predicts too many small cavities and misses some large cavities). However, we reiterate that when evaluating the numerous images comprising our complete test set, our model shows good assessments of material swelling on average.

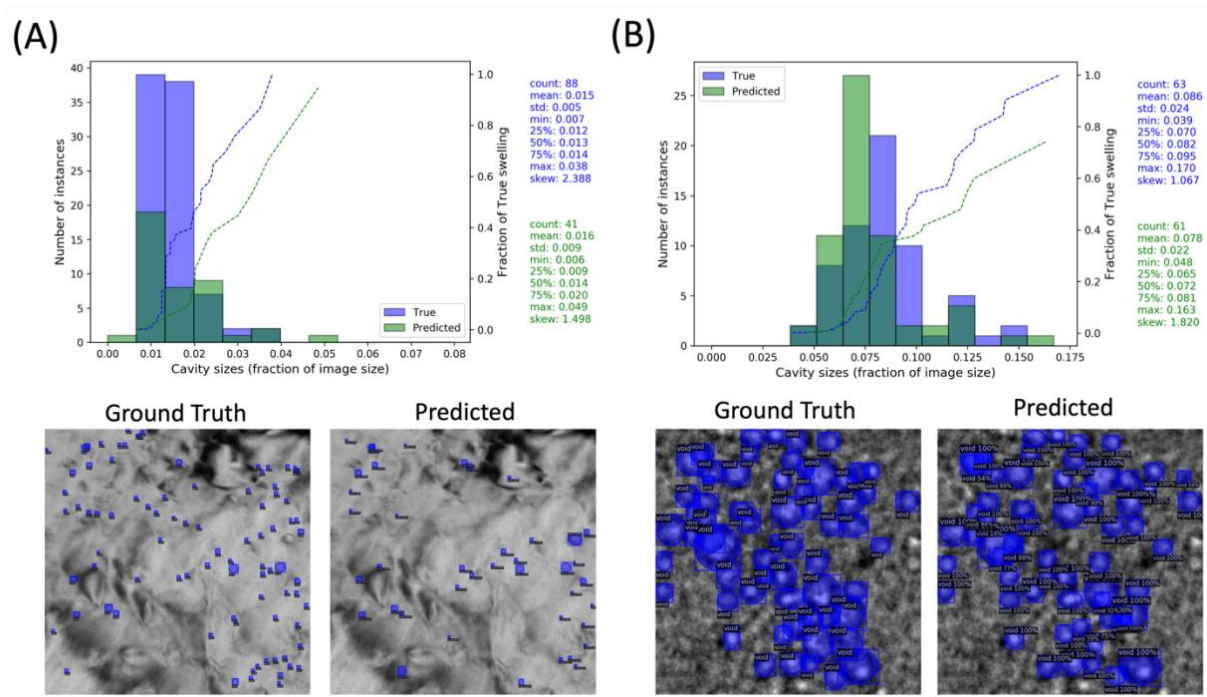

**Figure 5:** Plots of the cavity size (given here as fraction of total image size) histogram distribution and cumulative swelling contributions for two cases: (A) NOME image named "10 59 K.png", which is an underfocused image where the model predicted a low F1 score of 0.50 and a swelling error of only 0.004 percent swelling. (B) CNL image named "02.jpg", which is an underfocused image where the model predicted a high F1 score of 0.90 and a large swelling error of 1.46 percent swelling. For the histograms in each panel, the blue and green bars denote the number of instances of true and predicted cavities in each size bin, respectively, and the dashed blue and green lines denote the fraction of swelling (normalized to the total true swelling value) one would obtain by calculating swelling using the respective cavity size distributions. The percentage error in predicted swelling for (A) and (B) corresponds to 5.4% and 27.0%, respectively.

## 3    Summary and Outlook

In this work, we used an end-to-end deep learning approach based on the Mask R-CNN model to detect and characterize nanoscale cavities in irradiated metal alloy TEM micrographs. We have assembled the largest database of labeled cavity images to date, which includes 400 images and >34k cavities, with a domain encompassing an array of alloy compositions and irradiation conditions. We evaluated the performance of our Mask R-CNN models using a set of canonical classification-based metrics (overall and per-image precision, recall, and F1 scores) and materials domain-specific metrics of cavity size, cavity density, and swelling assessments. Given the importance of accurately characterizing swelling in irradiated alloys for their use as materials in nuclear reactor components, we particularly emphasized assessments of material swelling. Our model provides material swelling assessments with an average (standard deviation) swelling mean absolute error based on random leave-out cross validation of 0.30 (0.03) percent swelling, demonstrating good assessment ability of swelling with sufficiently small error to provide useful insight for new alloy design. We investigated the source of our swelling errors in greater detail, with three related findings of interest:

(1) The model can occasionally have poor assessments of cavity density, but these poor density assessments always coincided (at least for the images evaluated here) with small swelling errors as the missed cavities were all small (e.g., cavities which span about 2% or less of the image size), indicating that poor cavity density assessments are not necessarily a worrisome sign for model performance.

(2) Canonical classification-based metrics can sometimes paint a misleading picture of how well a model may perform for a specific materials-domain application. For example, we analyzed two extreme cases of test images with low (high) F1 scores which, in turn, ended up displaying very low (high) swelling errors, indicating that, like with point (1) above, missing many cavities is not necessarily an issue, assuming they are small.

(3) Directly related with the above points, which is given that swelling scales with the cube of cavity sizes, it is essential to capture the sizes of large cavities accurately. While this is obvious from inspection of Eq. 1, we showed how this effect can manifest in practice, where even test images with small average cavity size errors may show larger-than-desired errors in swelling, where in some cases errors in the full cavity size distribution, at least as it relates to accurately assessing swelling, are mainly the result of errors in cavity sizes of about 15 nm or larger.

Although the present results are very promising, the inability to reliably assess new types of cavity data, the errors on small cavity detection, and the swelling errors introduced for some large cavities are all still concerns. Some or all of these issues may be overcome with more data, but obtaining and annotating new TEM images of irradiated samples is very time-consuming, particularly if one also needs to conduct the irradiation experiments before imaging. We believe a potentially fruitful area of future research is to include synthetic training data, which can augment existing experimental databases to expand the model training domain to include different size distributions, focusing and imaging conditions, and noise levels to improve model training. One avenue for creating synthetic data is to use generative models such as Generative Adversarial Networks (GANs). However, the main downside of using GANs is their reliance on an initial set of training images of cavities. A different method that doesn't rely on an initial seed of training data is a physics-based simulation of cavities. Our initial work in this space combined simulated cavities onto experimental images containing real cavities to improve object detection model training,[32] and work is ongoing to address challenges of how to best integrate synthetic cavities with background TEM images and comprehensively evaluate object detection model performance with the addition of synthetic cavity data.

To encourage future studies of object detection and quantification in this space, we have made our full database of images and their associated ground truth annotations publicly available (see **Data and Code Availability** section). In addition, we have provided a Python notebook tailored for running on the free GPU resources provided on Google Colab, to easily provide inference and basic analysis of material swelling on user-provided test images. Finally, our model is also hosted on DLHub,[33] which is part of the Foundry for data, models and science.[34] This infrastructure enables inference on new images using only two lines of python code. We have also included a notebook which can be used to call our model from Foundry (see **Data and Code Availability** section). The Mask R-CNN model used for this tool was trained on the complete CNL+NOME database of 400 images to create the most accurate present model for detecting cavities on new images. Provided a new test image, the notebook saves the image with the model-specified cavity segmentations overlaid, together with a spreadsheet containing the bounding box, segmentation, and calculated size of each cavity in the image, along with the computed cavity density and swelling. We hope that tools such as these assist researchers and new users alike in the short term by creating a reduced barrier to using object detection tools. In the longer term, we hope to facilitate the generation of a broader community base of standardized (experimental and synthetic) image data and associated object detection models for the goal of creating state-of-the-art models able to accurately detect cavities and quantify vital materials properties such as swelling for a range of alloy compositions, irradiation doses, and imaging conditions.

## 4 Data and Methods

In this work, two datasets were used to train and test the performance of our Mask R-CNN object detection model. Both datasets consist of TEM images of irradiated metal alloys. Objects of interest for detection and quantification are cavities, which generally appear as spherical and faceted shapes in the micrsostructure with contrast consistent with a region devoid of matrix material). The first dataset consists of bright-field TEM micrographs obtained and labeled by the Canadian Nuclear Laboratory (CNL), which we refer to as the CNL dataset throughout this work. The images were obtained from reactor spacer springs of commercial nuclear reactors in the Canada Deuterium Uranium (CANDU) reactor fleet,[35] and consist of both overfocused and underfocused images of cavities in Inconel X-750 Ni alloys which have undergone neutron irradiation. The reactor spacer springs used to obtain the CNL images were in reactor service for 14 years, with a damage dose of 30 displacements per atom (dpa). Additional details of the sample preparation, TEM imaging, and cavity annotation are described in the work of Anderson et al.[31] Summary information of the number of overfocused and underfocused images, and the corresponding number of overfocused and underfocused cavities for the CNL dataset is summarized in **SI Note 4**. We note here that in the work of Anderson et al., it is stated that a total of 253 images comprise the database, where 230 images were used for training and 23 were reserved for testing their Faster R-CNN model. However, from the publicly available data linked in their paper, the available training set consists of 224 images and the testing set contains 19 images (243 total images). Further, when inspecting the provided annotations for all images, it was found that for 5 images, the annotations did not coincide with the cavities present on the image. Rather than re-annotating these images, we simply removed them from our present CNL database used in this work, yielding a total of 238 images. (Note the names of the 5 removed images are: 59_01.jpg, 59_02.jpg, 59_03.jpg, 59_04.jpg, 63_01.jpg). While 68% of the present CNL database consists of underfocused images, a large majority (about 83%) of the cavities are underfocused, resulting in a class imbalance where the database is significantly biased toward underfocused cavities.

The second dataset consists of TEM micrographs obtained and labeled by us as part of the Nuclear Oriented Materials & Examination (NOME) Laboratory at the University of Michigan, which we refer to as the NOME dataset throughout this work. These images were obtained through a wide variety of collaborations and professional contacts within the field. They consist of both overfocused and underfocused images. The materials compositions covered by these images are highly varied, including samples comprised of CW-316, T91, HT9, and 800H steel alloys. The irradiation undergone by each sample was also highly diverse and includes both damage received by light and heavy-ion as well as neutron bombardment, with total doses of up to 100 dpa. For annotating these images, a team of undergraduate student researchers were first trained by a domain expert to label images by practicing on several pre-labeled images not part of the NOME database. Feedback on their labeling was provided until results approximated those obtained by expert researchers. Once trained, the undergraduate team labeled the entire NOME database. The labels of each NOME database were corrected by a graduate student researcher (Matthew Lynch) and checked by a post-doctoral researcher (Priyam Patki) to form the final set of annotations. All labeling was done using the VGG Image Annotator (VIA) web tool.[36] The labeled NOME database comprises 162 images, as detailed in **SI Note 4**. Like the CNL database, the NOME database is significantly biased toward underfocused cavities, with about 75% of the total cavities coming from underfocused images. In order to assess different aspects of the model, 7 different splits of our combined CNL+NOME dataset were used to train and test the ability of our Mask R-CNN models to detect and quantify cavities, as detailed in **SI Note 4**. We note here that all of the images and annotations for the CNL and NOME datasets have been made publicly available on Figshare (see **Data and Code Availability** section).

We use the Mask R-CNN object detection model to detect and quantify cavities in this work, as implemented in the Detectron2 package (PyTorch backend). The Detectron2 package was developed by the Facebook AI Research (FAIR) team.[37] Detectron2 is freely available and enables the implementation of many object detection models, such as Faster R-CNN,[19] Mask R-CNN,[20] and Cascade R-CNN.[38] These object detection models have been pre-trained on either the ImageNet[17] or Microsoft COCO[18] (Common Objects in Context) image databases, enabling the use of the transfer learning technique. When using transfer learning, the model backbone weights are frozen to those obtained from the previous ImageNet or Microsoft COCO image training, save for a small number of terminal layers (2 throughout this work). The Mask R-CNN input configuration was the same as that used in our previous work of detecting and quantifying dislocation loops and black dot defects in FeCrAl alloys,[25] except here we adjusted the candidate anchor box sizes to be 4, 8, 16, 32, 64, 128, and 256 pixels to enable the model to better detect small cavities. We note here that input files in the Detectron2 package typically use candidate anchor box sizes that are powers of 2, so we follow that practice and also include the small anchor box sizes of 4 and 8 pixels in an effort to better detect small cavities, as some of the images examined in this work contain cavities that are on this length scale.

This work evaluates our model using both *classification-centric* and *materials property-centric* metrics. For our classification metrics, we focus on the model precision (P), recall (R), and F1 (harmonic mean of precision and recall) scores. Since we have only a single prediction category (i.e., cavities), the precision is calculated by dividing the number of found defects by the number of predicted defects, and the recall is calculated by dividing the number of found defects by the number of true defects. We evaluate P, R and F1 scores both on a per-image basis, from which we can obtain average per-image P, R and F1 scores, and we evaluate the so-called overall P, R and F1 scores, which is a single calculation using the total numbers of true, predicted and found cavities for the entire test set. For the materials property metrics, we calculate size distributions of predicted cavities for every test image, but focus our evaluation on comparing the true vs. predicted per-image average cavity size, true vs. predicted per-image cavity density (obtained by counting the number of true and predicted cavities in an image and dividing by the image area), and true vs. predicted per-image swelling. The swelling $\frac{\Delta V}{V}$ of an image (expressed as percent swelling) is calculated following the work of Jiao et al.[9]:

$$\frac{\Delta V}{V} = 100 \times \frac{\frac{\pi}{6}\sum_{i=1}^{N} d_i^3}{A\delta - \frac{\pi}{6}\sum_{i=1}^{N} d_i^3}, \qquad (1)$$

where $A$ is the area of the image, $\delta$ is the image thickness, $d_i$ is the cavity diameter, and $N$ is the number of cavities in the image. Due to the lack of per-image thickness data, we have assumed that every image has a thickness of 100 nm. The cavity diameter is calculated as twice its radius, where the cavity radius is defined as the square root of the product of the minimum and maximum distances from the center of the cavity mask.

When evaluating the performance of object detection models like Mask R-CNN, there are two key hyperparameters to choose from, namely the intersection-over-union (IoU) threshold value, and the objectness score. The IoU threshold determines the cutoff between ground truth and predicted bounding boxes to determine when a cavity can be considered found in the correct position, and the objectness score is a measure of the model confidence that a predicted region corresponds to a cavity, and thus impacts the total number of predicted cavities. The method to match the true and predicted cavities based on IoU is the same as that employed in our previous work.[25] We provide a brief summary of this approach here. When evaluating an image, there is a list of true defect masks and predicted defect masks. To decide whether a defect has been found in the correct location, the IoU of every predicted defect is calculated for each true defect, and the defect with the highest IoU score is considered the best possible match. The IoU values are calculated using the bounding boxes obtained from the region proposal network. If this computed IoU score is above the designated threshold, this predicted defect is considered to be found. Each true defect can only be found one time, so if multiple predicted defects are found to pass the IoU threshold with a particular true defect, the predicted defect with the highest IoU score is considered the found defect, and the other defect(s) would then be considered false positives. The hyperparameters will be determined using the CNL+NOME initial split by evaluating the overall F1 score as a function of IoU threshold and objectness score, and by evaluating the error in predicted swelling as a function of objectness score (see **Figure S5** in **SI Note 5**). This data split was chosen for hyperparameter optimization as it contains a representative and random subset of the full CNL+NOME image dataset examined in this work.

**Data and Code Availability**

The datasets generated during and/or analyzed during the current study are available on Figshare (https://doi.org/10.6084/m9.figshare.20063117). The trained model on the full database of all CNL and NOME images, a Google Colab notebook and associated python scripts to make predictions on new images and save the associated data is also available on Figshare (https://doi.org/10.6084/m9.figshare.20063117). In addition, we have hosted the final trained model on DLHub, which is part of the Foundry for data, models and science. A notebook to use the hosted model on Foundry is also provided in the above Figshare repository. A small subset of the images (≈3%) are omitted from the public database due to protected rights of these images. Access to the omitted images and corresponding labels can be obtained through request with the corresponding author.

**Acknowledgments**

This work was funded by the Electric Power Research Institute (EPRI) under award number 10012138. This work used the Extreme Science and Engineering Discovery Environment (XSEDE), which is supported by National Science Foundation grant number ACI-1548562. Specifically, it used the Bridges-2 system through allocation TG-DMR090023, which is supported by NSF award number ACI-1928147, at the Pittsburgh Supercomputing Center (PSC).[39]

**Competing Interests**

The authors declare no competing interests.

**References**

[1]   A. Seeger, J. Diehl, S. Mader, H. Rebstock, Work-hardening and work-softening of face-centred cubic metal crystals, Philos. Mag. 2 (1957) 323–350. doi:10.1080/14786435708243823.
[2]   C. Zheng, E.R. Reese, K.G. Field, E. Marquis, S.A. Maloy, D. Kaoumi, Microstructure response of ferritic/martensitic steel HT9 after neutron irradiation: effect of dose, J. Nucl. Mater. 523 (2019) 421–433. doi:10.1016/j.jnucmat.2019.06.019.
[3]   H.K. Zhang, Z. Yao, C. Judge, M. Griffiths, Microstructural evolution of CANDU spacer material Inconel X-750 under in situ ion irradiation, J. Nucl. Mater. 443 (2013) 49–58. doi:10.1016/j.jnucmat.2013.06.034.
[4]   D.L. Porter, F.A. Garner, Irradiation creep and embrittlement behavior of AISI 316 stainless steel at very high neutron fluences, J. Nucl. Mater. 159 (1988) 114–121. doi:10.1016/0022-3115(88)90089-X.
[5]   F.A. Garner, Recent insights on the swelling and creep of irradiated austenitic alloys, J. Nucl. Mater. 122 (1984) 459–471. doi:10.1016/0022-3115(84)90641-X.
[6]   E. Snoeck, J. Majimel, M.O. Ruault, M.J. Hÿtch, Characterization of helium bubble size

and faceting by electron holography, J. Appl. Phys. 100 (2006). doi:10.1063/1.2216791.

[7] D. Morgan, G. Pilania, A. Couet, B.P. Uberuaga, C. Sun, J. Li, Machine learning in nuclear materials research, Curr. Opin. Solid State Mater. Sci. 26 (2022) 100975. doi:10.1016/j.cossms.2021.100975.

[8] B. V. Cockeram, R.W. Smith, N. Hashimoto, L.L. Snead, The swelling, microstructure, and hardening of wrought LCAC, TZM, and ODS molybdenum following neutron irradiation, J. Nucl. Mater. 418 (2011) 121–136. doi:10.1016/j.jnucmat.2011.05.055.

[9] Z. Jiao, S. Taller, K. Field, G. Yeli, M.P. Moody, G.S. Was, Microstructure evolution of T91 irradiated in the BOR60 fast reactor, J. Nucl. Mater. 504 (2018) 122–134. doi:10.1016/j.jnucmat.2018.03.024.

[10] C.A. Schneider, W.S. Rasband, K.W. Eliceiri, NIH Image to ImageJ: 25 years of image analysis, Nat. Methods. 9 (2012) 671–675. doi:10.1038/nmeth.2089.

[11] L.A. Giannuzzi, J.L. Drown, S.R. Brown, R.B. Irwin, F.A. Stevie, Applications of the FIB lift-out technique for TEM specimen preparation, Microsc. Res. Tech. 41 (1998) 285–290. doi:10.1002/(SICI)1097-0029(19980515)41:4<285::AID-JEMT1>3.0.CO;2-Q.

[12] A. Schemer-Kohrn, M.B. Toloczko, Y. Zhu, J. Wang, D.J. Edwards, Removal of FIB Damage using Flash Electropolishing for Artifact-free TEM Foils, Microsc. Microanal. 25 (2019) 1606–1607. doi:10.1017/S1431927619008766.

[13] S.R. Spurgeon, C. Ophus, L. Jones, A. Petford-Long, S. V. Kalinin, M.J. Olszta, R.E. Dunin-Borkowski, N. Salmon, K. Hattar, W.C.D. Yang, R. Sharma, Y. Du, A. Chiaramonti, H. Zheng, E.C. Buck, L. Kovarik, R.L. Penn, D. Li, X. Zhang, M. Murayama, M.L. Taheri, Towards data-driven next-generation transmission electron microscopy, Nat. Mater. 20 (2021) 274–279. doi:10.1038/s41563-020-00833-z.

[14] Y. Jiang, Z. Chen, Y. Han, P. Deb, H. Gao, S. Xie, P. Purohit, M.W. Tate, J. Park, S.M. Gruner, V. Elser, D.A. Muller, Electron ptychography of 2D materials to deep sub-ångström resolution, Nature. 559 (2018) 343–349. doi:10.1038/s41586-018-0298-5.

[15] D. Chatterjee, J. Wei, A. kvit, B. Bammes, B. Levin, R. Bilhorn, P. Voyles, An Ultrafast Direct Electron Camera for 4D STEM, Microsc. Microanal. 27 (2021) 1004–1006. doi:10.1017/s1431927621003809.

[16] C. Ophus, Four-Dimensional Scanning Transmission Electron Microscopy (4D-STEM): From Scanning Nanodiffraction to Ptychography and Beyond, Microsc. Microanal. (2019) 563–582. doi:10.1017/S1431927619000497.

[17] J. Deng, W. Dong, R. Socher, L.-J. Li, K. Li, L. Fei-Fei, ImageNet: A Large-Scale Hierarchical Image Database, 2009 IEEE Conf. Comput. Vis. Pattern Recognit. (2009).

[18] T.-Y. Lin, M. Maire, S. Belongie, J. Hays, P. Perona, D. Ramanan, P. Dollar, C.L. Zitnick, Microsoft COCO: Common Objects in Context, Eur. Conf. Comput. Vis. (2014) 740–755.

[19] S. Ren, K. He, R. Girshick, J. Sun, Faster R-CNN: Towards Real-Time Object Detection with Region Proposal Networks, IEEE Trans. Pattern Anal. Mach. Intell. 39 (2017) 1137–1149. doi:10.1109/TPAMI.2016.2577031.

[20] K. He, G. Gkioxari, P. Dollar, R. Girshick, Mask R-CNN, Int. Conf. Comput. Vis. (2017).

[21] L. Liu, W. Ouyang, X. Wang, P. Fieguth, J. Chen, X. Liu, M. Pietikäinen, Deep Learning for Generic Object Detection: A Survey, Int. J. Comput. Vis. 128 (2020) 261–318. doi:10.1007/s11263-019-01247-4.

[22] Z.Q. Zhao, P. Zheng, S.T. Xu, X. Wu, Object Detection with Deep Learning: A Review, IEEE

Trans. Neural Networks Learn. Syst. 30 (2019) 3212–3232. doi:10.1109/TNNLS.2018.2876865.

[23] R. Jacobs, Deep learning object detection in materials science : Current state and future directions, Comput. Mater. Sci. 211 (2022) 111527. doi:10.1016/j.commatsci.2022.111527.

[24] G. Roberts, S.Y. Haile, R. Sainju, D.J. Edwards, B. Hutchinson, Y. Zhu, Deep Learning for Semantic Segmentation of Defects in Advanced STEM Images of Steels, Sci. Rep. 9 (2019). doi:10.1038/s41598-019-49105-0.

[25] R. Jacobs, M. Shen, Y. Liu, W. Hao, X. Li, R. He, J.R.C. Greaves, D. Wang, Z. Xie, Z. Huang, C. Wang, K.G. Field, D. Morgan, Performance and limitations of deep learning semantic segmentation of multiple defects in transmission electron micrographs, Cell Reports Phys. Sci. (2022) 100876. doi:10.1016/j.xcrp.2022.100876.

[26] M. Shen, G. Li, D. Wu, Y. Liu, J. Greaves, W. Hao, N.J. Krakauer, L. Krudy, J. Perez, V. Srrenivasan, B. Sanchez, O. Torres-Velazquez, W. Li, K.G. Field, D. Morgan, Multi Defect Detection and Analysis of Electron Microscopy Images with Deep Learning, Comput. Mater. Sci. 199 (2021) 110576. https://doi.org/10.1016/j.commatsci.2021.110576.

[27] R. Cohn, I. Anderson, T. Prost, J. Tiarks, E. White, E. Holm, Instance Segmentation for Direct Measurements of Satellites in Metal Powders and Automated Microstructural Characterization from Image Data, Jom. 73 (2021) 2159–2172. doi:10.1007/s11837-021-04713-y.

[28] C.K. Groschner, C. Choi, M.C. Scott, Machine Learning Pipeline for Segmentation and Defect Identification from High-Resolution Transmission Electron Microscopy Data, Microsc. Microanal. 27 (2021) 549–556. doi:10.1017/S1431927621000386.

[29] M. Ziatdinov, O. Dyck, A. Maksov, X. Li, X. Sang, K. Xiao, R.R. Unocic, R. Vasudevan, S. Jesse, S. V. Kalinin, Deep Learning of Atomically Resolved Scanning Transmission Electron Microscopy Images: Chemical Identification and Tracking Local Transformations, ACS Nano. 11 (2017) 12742–12752. doi:10.1021/acsnano.7b07504.

[30] M. Ge, H.L. Xin, Deep Learning Based Atom Segmentation and Noise and Missing-Wedge Reduction for Electron Tomography, Microsc. Microanal. 24 (2018) 504–505. doi:10.1017/s143192761800301x.

[31] C.M. Anderson, J. Klein, H. Rajakumar, C.D. Judge, L.K. Béland, Automated Detection of Helium Bubbles in Irradiated X-750, Ultramicroscopy. 217 (2020) 113068. doi:10.1016/j.ultramic.2020.113068.

[32] K.G. Field, R. Jacobs, M. Shen, M. Lynch, P. Patki, C. Field, D. Morgan, Development and Deployment of Automated Machine Learning Detection in Electron Micrcopy Experiments, Microsc. Microanal. 27 (2021) 2136–2137. doi:10.1017/s1431927621007704.

[33] R. Chard, Z. Li, K. Chard, L. Ward, Y. Babuji, A. Woodard, S. Tuecke, B. Blaiszik, M.J. Franklin, I. Foster, DLHub: Model and data serving for science, Proc. - 2019 IEEE 33rd Int. Parallel Distrib. Process. Symp. IPDPS 2019. (2019) 283–292. doi:10.1109/IPDPS.2019.00038.

[34] U. of Chicago, U. of Wisconsin-Madison, Foundry Materials Informatics Environment:, (2021). https://ai-materials-and-chemistry.gitbook.io/foundry/v/docs/.

[35] H.K. Zhang, Z. Yao, G. Morin, M. Griffiths, TEM characterization of in-reactor neutron

irradiated CANDU spacer material Inconel X-750, J. Nucl. Mater. 451 (2014) 88–96. doi:10.1016/j.jnucmat.2014.03.043.

[36] A. Dutta, A. Gupta, A. Zisserman, VGG Image Annotator (VIA), (2020). https://www.robots.ox.ac.uk/~vgg/software/via/ (2020).

[37] Y. Wu, A. Kirillov, F. Massa, W.-Y. Lo, R. Girshick, Detectron2, (2019). https://github.com/facebookresearch/detectron2 (2019).

[38] Z. Cai, N. Vasconcelos, Cascade R-CNN: Delving into High Quality Object Detection, Proc. IEEE Conf. Comput. Vis. Pattern Recognit. (2018) 6154–6162.

[39] N. Towns, J., Cockerill, T., Dahan, M., Foster, I., Gaither, K., Grimshaw, A., Hazlewood, V., Lathrop, S., Lifka, D., Peterson, G., Roskies, R., Scott, J., Wilkins-Diehr, XSEDE: Accelerating Scientific Discovery, Comput. Sci. Eng. 120 (2014) 4–5.

# Supplementary Information for

## Materials Swelling Revealed Through Automated Semantic Segmentation of Cavities in Electron Microscopy Images

Ryan Jacobs[1], Priyam Patki[2], Matthew Lynch[2], Steven Chen[2], Dane Morgan[1], Kevin G. Field[2]

[1]Department of Materials Science and Engineering, University of Wisconsin-Madison, Madison, Wisconsin, 53706, USA

[2]Nuclear Engineering and Radiological Sciences, University of Michigan - Ann Arbor, Michigan, 48109 USA

[†]Corresponding author email: rjacobs3@wisc.edu

### SI Note 1: Additional results of model swelling predictions

Regarding the cavity sizes in **Figure S6A**, we see that the model can predict the average per-image cavity size with high accuracy, with an average (standard deviation) MAE of just 1.02 (0.14) nm, which corresponds to an average (standard deviation) MAPE value of 8.94% (0.84%) error in cavity size. This level of defect size accuracy is consistent with our previous work employing the Mask R-CNN model to detect and quantify dislocation loops and black spot defects in FeCrAl alloys, which showed average (standard deviation) defect percent errors of 7.3% (3.8%) also from random cross validation.[2] While this scale of error can generally be considered small, we show in **Section 2.3** of the main text that the errors in swelling are still mainly due to model errors in the full cavity size distribution, particularly for images containing large (> 15 nm) cavities. Finally, the predictions of per-image cavity density are presented in **Figure S6B**. From the parity plot fit statistics standpoint, our model has the highest errors in predicting cavity density. This is particularly true for images with high cavity densities (> 20 ×10$^4$ nm$^{-2}$), where the model has a clear bias to lower values. We show in **Section 2.3** of the main text that the significant errors for images with high cavity densities result from the model having difficulty identifying many small (< 5 nm, or about 2% of the image dimension) cavities. **Table S1** contains a summary of the key classification metrics and materials property metrics for each split, together with the average and standard deviation across all five splits.

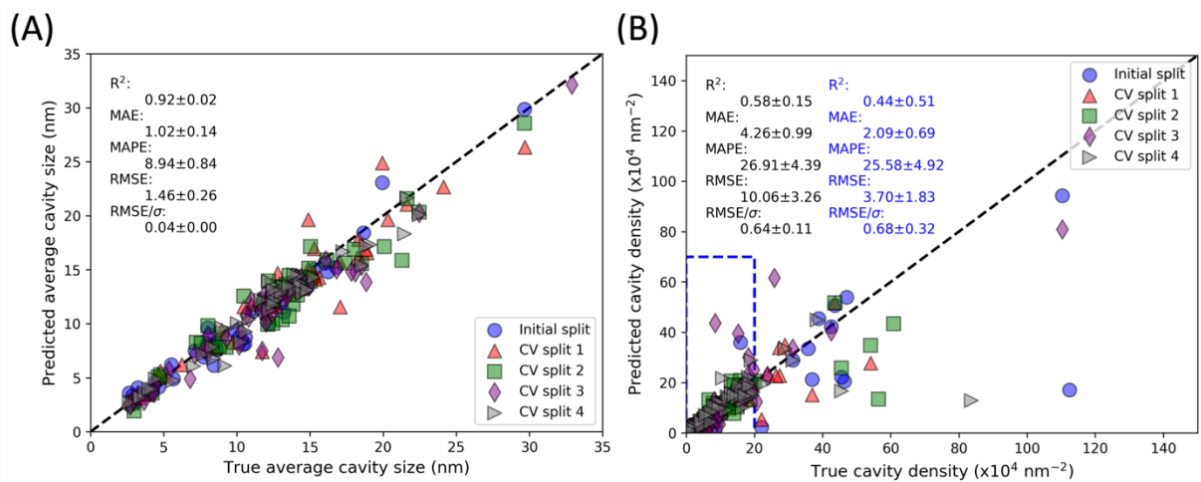

**Figure S6:** Parity plot of true and predicted (A) average cavity size, and (B) cavity density. Each data point represents one test image. The different symbols correspond to different cross validation train/test splits. The fit statistics in black text denote the average +/- standard deviation across all five splits for each metric. In (B), the fit statistics in blue text denotes the average +/- standard deviation across all five splits for test images with true cavity density equal or less than 20 x $10^4$ nm$^{-2}$.

**Table S1:** Summary of classification and material property metrics for five splits of random cross validation using the combined CNL+NOME dataset.

| Data split | Overall statistic | | | Average per-image statistic | | | Defect size error (nm) (percent error) | Defect density error (x$10^4$ nm$^{-2}$) (percent error) | Defect swelling error (%) (percent error) |
|---|---|---|---|---|---|---|---|---|---|
| | P | R | F1 | P | R | F1 | | | |
| Initial split | 0.74 | 0.62 | 0.68 | 0.78 | 0.65 | 0.69 | 0.82 (9.73) | 5.94 (25.96) | 0.26 (32.74) |
| CV split 1 | 0.74 | 0.67 | 0.70 | 0.75 | 0.71 | 0.71 | 1.18 (9.10) | 2.92 (24.00) | 0.35 (20.07) |
| CV split 2 | 0.70 | 0.62 | 0.66 | 0.73 | 0.71 | 0.70 | 1.10 (9.16) | 3.82 (25.25) | 0.33 (24.78) |
| CV split 3 | 0.72 | 0.62 | 0.67 | 0.70 | 0.71 | 0.68 | 1.09 (9.38) | 4.53 (35.54) | 0.32 (24.44) |
| CV split 4 | 0.75 | 0.71 | 0.71 | 0.75 | 0.74 | 0.73 | 0.88 (7.32) | 4.12 (23.80) | 0.27 (19.12) |
| *Average over splits* | *0.73* | *0.65* | *0.69* | *0.74* | *0.70* | *0.70* | *1.02 (8.94)* | *4.26 (26.91)* | *0.30 (24.23)* |
| *Standard deviation over splits* | *0.02* | *0.04* | *0.02* | *0.03* | *0.03* | *0.02* | *0.14 (0.84)* | *0.99 (4.39)* | *0.03 (4.82)* |

**SI Note 2: Additional results of leave out group cross validation tests**

Regarding **Figure 3B** in the main text, from the values in **Table S2**, we can see that the model trained and tested on the NOME data performs better on overfocused images than underfocused images, where the swelling MAE values on underfocused (overfocused) images are 0.20 (0.07) percent swelling, respectively. This behavior is opposite to what was observed for the model trained and tested on CNL data. From further inspecting the statistics in **Table S2**, we surmise that the lower swelling error for overfocused images for the NOME model is the result of the lower cavity size errors of just 0.56 nm (vs. 1.02 nm for underfocus), even though the model has worse cavity density errors of 13.84 x$10^4$ nm$^{-2}$ for overfocused images (vs. 3.52 x$10^4$ nm$^{-2}$ for underfocused images). This result makes sense, given that the swelling scales with the cube of the cavity sizes but only linearly with cavity density (see Eq. 1 in **Section 4** the main text). Therefore, it is more critical to obtain accurate cavity sizes for accurate swelling predictions than accurate cavity densities. This interplay of model errors of cavity size, density and swelling is discussed more in **Section 2.3** of the main text.

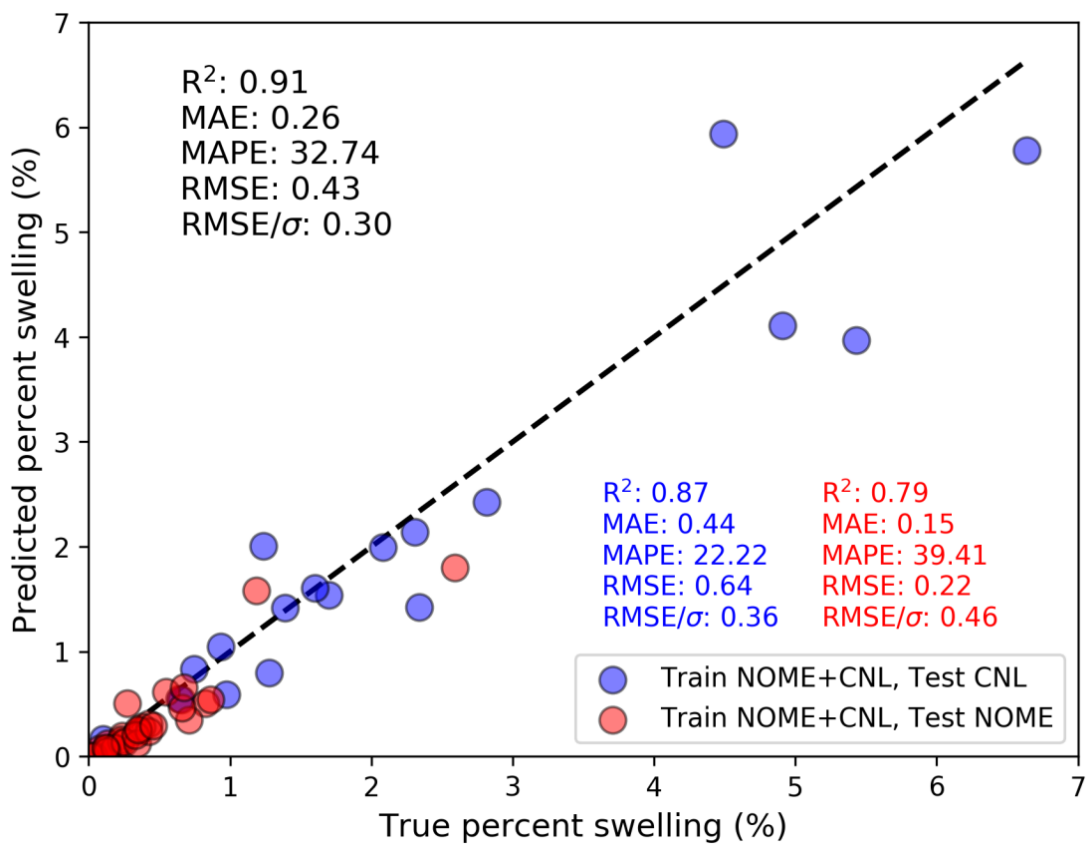

**Figure S7:** Additional parity plot assessing Mask R-CNN per-image performance of predicting materials swelling. CNL+NOME initial split, with model trained on CNL+NOME and tested on CNL+NOME, where CNL images are shown as blue points and NOME images as red points.

**Table S2:** Summary of materials property metrics of per-image defect density, defect size, and swelling error for Mask R-CNN models fit to the data splits shown in **Figure 3** of the main text and **Figure S7**. The quoted error values are all MAE values, and the quoted percent errors are all MAPE values.

| Split name | Train | Test | Figure | Defect size error (nm) (percent error) | Defect density error (x10$^4$ nm$^{-2}$) (percent error) | Defect swelling error (%) (percent error) |
|---|---|---|---|---|---|---|
| CNL initial split | CNL | CNL | Figure 3A | 0.68 (5.61) Over: 0.81 (5.87) Under: 0.60 (5.46) | 2.77 (16.39) Over: 1.67 (13.83) Under: 3.41 (17.89) | 0.40 (19.68) Over: 0.53 (19.08) Under: 0.33 (20.04) |
| CNL initial split | CNL | NOME | Figure 3A | 4.49 (55.41) Over: 3.45 (51.43) | 14.78 (87.95) Over: 25.19 (176.0) | 0.68 (215.29) Over: 0.20 (121.28) |

| | | | | Under: 5.08 (57.71) | Under: 8.75 (36.97) | Under: 0.96 (269.72) |
|---|---|---|---|---|---|---|
| NOME initial split | NOME | NOME | Figure 3B | 0.85 (11.92) Over: 0.56 (11.90) Under: 1.02 (11.93) | 7.30 (28.88) Over: 13.84 (47.79) Under: 3.52 (17.93) | 0.15 (37.97) Over: 0.07 (46.22) Under: 0.20 (33.20) |
| NOME initial split | NOME | CNL | Figure 3B | 6.09 (43.09) Over: 8.09 (55.59) Under: 4.93 (35.80) | 12.23 (84.45) Over: 7.96 (71.84) Under: 14.73 (91.81) | 1.98 (76.25) Over: 2.47 (83.08) Under: 1.70 (72.26) |
| CNL+NOME initial split | CNL+NOME | CNL+NOME | Figure S2 | 0.82 (9.73) Over: 0.64 (9.00) Under: 0.93 (10.15) | 5.94 (25.96) Over: 8.82 (34.38) Under: 4.27 (21.07) | 0.26 (32.74) Over: 0.26 (39.13) Under: 0.27 (29.03) |

Regarding model performance on classification-centric metrics, all models perform better at detecting underfocused cavities than overfocused cavities, where, for example, the CNL+NOME model shows overall (average per-image) F1 scores of 0.72 (0.73) for underfocused images and 0.54 (0.60) for overfocused images, respectively. In addition, when considering underfocused and overfocused images together, the CNL+NOME model shows overall (average per-image) F1 scores of 0.68 (0.69), which are nearly identical to scores of 0.68 (0.73) for the model trained and tested solely on CNL data and to scores of 0.68 (0.66) for the model trained and tested solely on NOME data. Further, we speculate that the model trained solely on NOME data may have a slightly larger domain of applicability than the model trained solely on CNL data. The NOME database contains images of materials from more alloy types and irradiation conditions than the CNL database. From the data in **Table S3**, the model trained on NOME and tested on CNL displays a better overall F1 score of 0.46 than the model trained on CNL and tested on NOME, which has an overall F1 score of 0.39. Overall, the results of **Figure 3** in the main text, **Figure S7** and **Table S3** demonstrate that it is preferable to simply train one model with training images from both datasets, as the model domain is widened without significant loss in classification or materials property metric performance. Finally, it is worth noting that comparisons can be made for the CNL initial split model that is trained and tested on the CNL data with previous findings from the work of Anderson et al.[1] Our model shows overall and average per-image F1 scores of 0.68 and 0.73, respectively, which are lower than the highest F1 score of 0.78 reported in Anderson et al. While it is not clear whether this F1 score of 0.78 reported by Anderson et al. represents an overall or average per-image F1 score, it is nonetheless 0.05-0.1 higher than the F1 scores we obtain here. We attribute this difference to being due to the different number of training images used here compared to Anderson et al. (219 vs. 230 in their work), and the different test set used to evaluate the model performance (19 images vs. 23 images in their work). In addition, different codebases and model types were used between this work and Anderson et al., who used a Tensorflow-based implementation of the Faster R-CNN model, while we use the Mask R-CNN model in Detectron2/Pytorch.

**Table S3:** Summary of classification metrics of per-image P, R, F1 scores and overall P, R and F1 scores for Mask R-CNN models fit to the data splits shown in **Figure 3** of the main text and **Figure S7**.

| Split name | Train | Test | Figure | Overall statistic (Row 1 = All, Row 2 = Overfocus, Row 3 = Underfocus) | | | Average per-image statistic (Row 1 = All, Row 2 = Overfocus, Row 3 = Underfocus) | | |
|---|---|---|---|---|---|---|---|---|---|
| | | | | P | R | F1 | P | R | F1 |
| CNL initial split | CNL | CNL | Figure 3A | 0.72 | 0.64 | 0.68 | 0.76 | 0.72 | 0.73 |
| | | | | 0.55 | 0.52 | 0.53 | 0.76 | 0.70 | 0.72 |
| | | | | 0.77 | 0.66 | 0.71 | 0.76 | 0.73 | 0.73 |
| CNL initial split | CNL | NOME | Figure 3A | 0.51 | 0.31 | 0.39 | 0.56 | 0.35 | 0.40 |
| | | | | 0.30 | 0.18 | 0.22 | 0.49 | 0.23 | 0.29 |
| | | | | 0.61 | 0.39 | 0.48 | 0.59 | 0.41 | 0.47 |
| NOME initial split | NOME | NOME | Figure 3B | 0.82 | 0.59 | 0.68 | 0.82 | 0.60 | 0.66 |
| | | | | 0.83 | 0.35 | 0.49 | 0.85 | 0.44 | 0.53 |
| | | | | 0.82 | 0.72 | 0.77 | 0.81 | 0.70 | 0.74 |
| NOME initial split | NOME | CNL | Figure 3B | 0.62 | 0.36 | 0.46 | 0.47 | 0.23 | 0.26 |
| | | | | 0.58 | 0.23 | 0.33 | 0.25 | 0.09 | 0.13 |
| | | | | 0.63 | 0.39 | 0.48 | 0.60 | 0.31 | 0.34 |
| CNL+NOME initial split | CNL+NOME | CNL+NOME | Figure S2 | 0.74 | 0.62 | 0.68 | 0.78 | 0.65 | 0.69 |
| | | | | 0.65 | 0.47 | 0.54 | 0.78 | 0.56 | 0.60 |
| | | | | 0.76 | 0.68 | 0.72 | 0.79 | 0.71 | 0.73 |

**SI Note 3: Additional discussion of swelling errors**

**Figure S8** contains a scatter plot of the true per-image average cavity size vs. the true per-image cavity density for all data splits considered. In **Figure S8**, the sizes of the points scale with the true material swelling. It is immediately evident in **Figure S8** that the images with the largest swelling are those with average cavity sizes of about 15 nm and greater, cavity areal densities of about 10 x $10^4$ nm$^{-2}$ and higher (note this corresponds to a volume density of about $10^{24}$ cavities/cm$^3$, assuming a thickness of 100 nm), and that the material swelling is much more sensitive to the cavity size than the density, consistent with intuition.

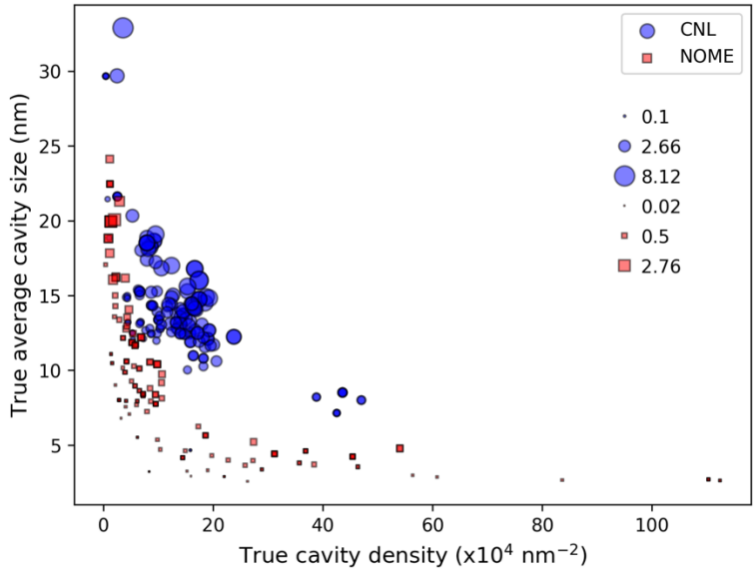

**Figure S8:** Relationship between the true per-image cavity size and the true per-image cavity density. Each data point represents one test image. Respectively, the blue circles and red squares denote CNL and NOME test images. The size of the data points scale with the true percent swelling.

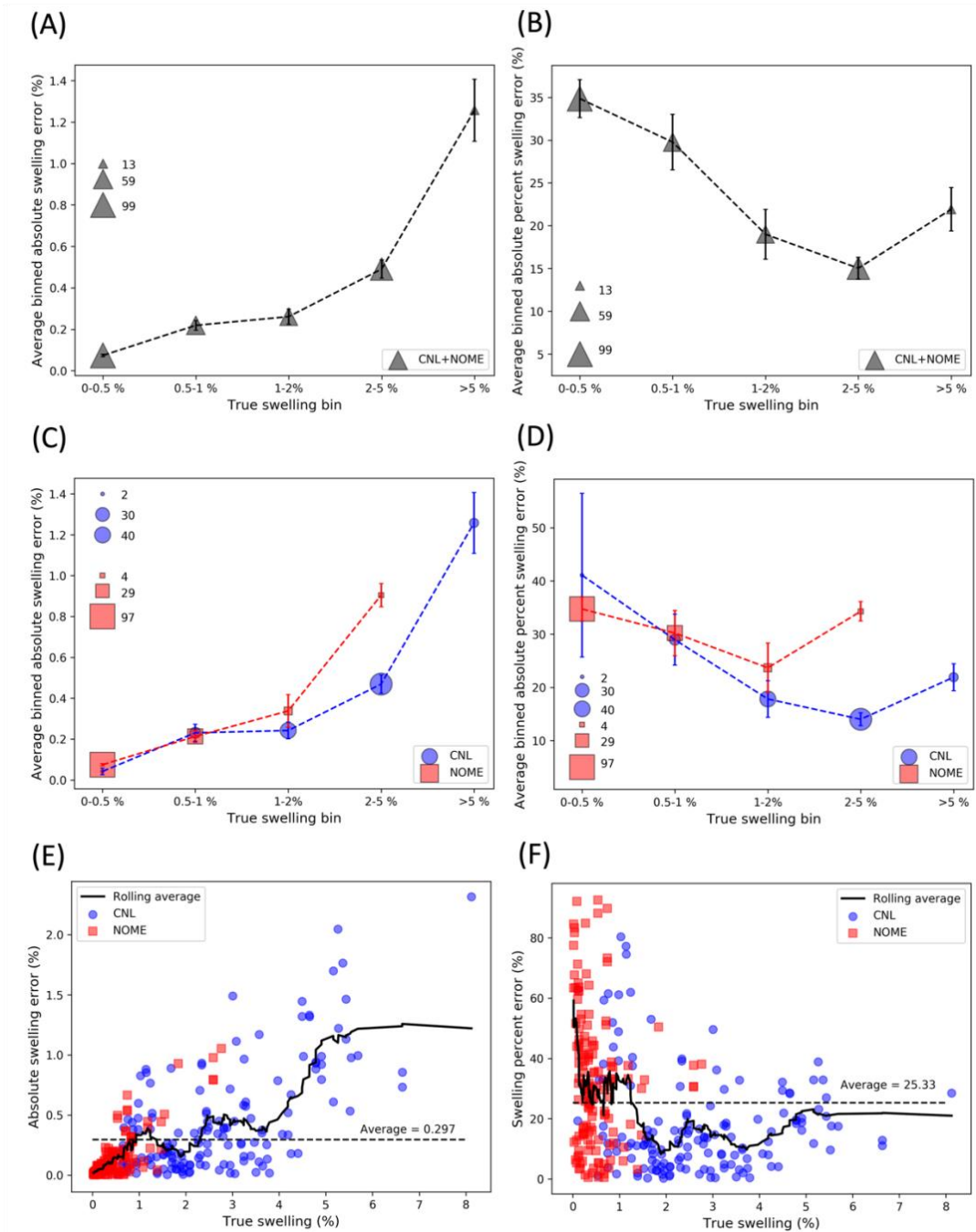

**Figure S9:** (A, C and E) Trend of model absolute error in material swelling as a function of true swelling. (B, D and F) Trend of model absolute percentage error in material swelling as a function of true swelling. In A-D, the x-axis represents binned values of true swelling. In C-F, the blue circles and red squares denote groups of CNL and NOME test images, respectively. The size of the points scales with the number of test images comprising the true swelling bin. The size legends denote the minimum, average, and maximum for the respective data trace. The error bars are the standard error in the mean.

**SI Note 4: Additional information on cavity datasets**

**Table S4:** Summary of image and cavity numbers of the CNL and NOME datasets. Note that some NOME images cannot strictly be classified as over or underfocus, so the total number of NOME images is larger than the sum of over and underfocus images.

| Dataset | Number of images | Number of cavities | Overfocus images | Underfocus images | Overfocus cavities | Underfocus cavities |
|---|---|---|---|---|---|---|
| CNL | 238 | 22,864 | 75 | 163 | 3,826 | 19,038 |
| NOME | 162 | 11,569 | 40 | 107 | 2,651 | 8,300 |

**Table S5** contains information detailing the data used to train and test each model and the total numbers of images and cavities used to train and test each model. In addition, a reference to the respective figure and table containing results of models trained and tested on a given data split are provided. Our data splits consist of models trained individually on the CNL (CNL initial split) or NOME (NOME initial split) datasets, which were then tested separately on the CNL and NOME test sets. The purpose of evaluating the Mask R-CNN model performance on the CNL initial split and NOME initial split was twofold: first, it provided information on model performance for models trained and tested on the same dataset (i.e., train on CNL, test on CNL, and train on NOME, test on NOME), and second, it provided a baseline assessment of the ability to use each model to predict properties of the dataset not used in training (i.e., train on CNL, test on NOME, and train on NOME, test on CNL), providing an assessment of the model applicability domain on test data which is markedly different from the training data in terms of material composition, irradiation condition, and typical cavity size.

After conducting tests of models trained individually on the CNL and NOME datasets, we trained a new model combining the same training and test images used in the CNL initial split and NOME initial split, corresponding to a new model called CNL+NOME initial split. This test aimed to assess whether a model could be developed that effectively extends the applicability domain of cavity detection and quantification to encompass both the CNL and NOME datasets, instead of relying on models trained on separate sub-domains of the data. This combined dataset model was then tested on the same test images of the CNL initial split and NOME initial split, enabling a comparison of the combined model performance to separately predict cavity properties of the CNL and NOME test data.

Finally, we further evaluate the model performance on combined CNL+NOME datasets by random cross-validation of the train and test image sets. The CNL+NOME initial split discussed above is one such split as it had effectively a random group of images pulled out for testing. We constructed an additional 4 random splits (for a total of 5 random splits), which we refer to as CNL+NOME CV split N (N=1-4) in

**Table S5** and throughout this work. The purpose of evaluating models with these different random splits of CNL+NOME data was to quantify an expected average and standard deviation in model predictive performance for the scenario where the test images are drawn approximately from the same domain as the training images.

**Table S5:** Summary of data splits used to train and test Mask R-CNN models in this work.

| Split name | Train | Test | Train images | Test images | Train cavities | Test cavities | Figure(s) containing results | Table(s) containing results |
|---|---|---|---|---|---|---|---|---|
| CNL initial split | CNL | CNL | 219 | 19 | 20,082 | 2782 | Figure 3A | Table S2, Table S3 |
| CNL initial split | CNL | NOME | 219 | 30 | 20,082 | 2154 | Figure 3A | Table S2, Table S3 |
| NOME initial split | NOME | NOME | 132 | 30 | 9415 | 2154 | Figure 3B | Table S2, Table S3 |
| NOME initial split | NOME | CNL | 132 | 19 | 9415 | 2782 | Figure 3B | Table S2, Table S33 |
| CNL+NOME initial split | CNL+NOME | CNL+NOME | 351 | 49 | 29,474 | 4936 | Figure 2 Figure 4, Figure S1, Figure S2, Figure S4 | Table S1, Table S2, Table S3 |
| CNL+NOME CV split 1 | CNL+NOME | CNL+NOME | 350 | 50 | 30,679 | 3756 | Figure 2 | Table S1 |
| CNL+NOME CV split 2 | CNL+NOME | CNL+NOME | 350 | 50 | 30,360 | 4075 | Figure 2 | Table S1 |
| CNL+NOME CV split 3 | CNL+NOME | CNL+NOME | 350 | 50 | 29,765 | 4670 | Figure 2 | Table S1 |
| CNL+NOME CV split 4 | CNL+NOME | CNL+NOME | 350 | 50 | 31,275 | 3160 | Figure 2 | Table S1 |

**SI Note 5: Hyperparameter determination**

The first step to evaluating the performance of our Mask R-CNN models for detecting and quantifying cavities is to choose the value of the IoU threshold and objectness score which maximizes the model performance. **Figure S10** contains two measures of model performance, both of which were obtained by using the CNL+NOME initial split data (see **Table S4** and **Table S5** in **SI Note 4** and **Section 4** of the main text for more information on data splits). **Figure S10A** contains a heatmap plotting the overall F1 score as a function of both IoU threshold and objectness score. From this assessment, we find that an IoU of 0.1 and objectness score of 0.1 result in the best performing model in terms of overall F1 score. **Figure S10B** contains a plot of the mean absolute error (MAE) and root mean squared error (RMSE) of the predicted percent swelling as a function of objectness score. We find that an objectness score of 0.1 results in the lowest swelling MAE of 0.26 percent swelling, and a swelling RMSE of 0.43 percent swelling, slightly higher than the lowest RMSE of 0.42 percent swelling for objectness scores of 0.3 and 0.5. Note that the swelling prediction is only a function of the objectness score and not the IoU threshold. The objectness score determines the total number of predicted cavities per image, thus affecting all of the performance statistics evaluated in this work, while the IoU threshold is used only for classifying when a predicted cavity can be matched with a corresponding true cavity, which in turn affects the P, R and F1 scores only. Based on the findings presented in **Figure S10** of highest overall F1 and lowest swelling MAE for IoU = 0.1 and objectness score = 0.1, for the remainder of this study we evaluate the performance of all models using this set of hyperparameter values. We note that a lower IoU threshold of 0.01 and objectness score of 0.01 resulted in worse performance with an overall F1 score of 0.658, a swelling MAE of 0.27 percent and swelling RMSE of 0.46 percent, indicating the IoU threshold and objectness values of 0.1 and 0.1, respectively, used here likely give optimum model performance, at least on this dataset. In addition, it is worth noting that this optimum IoU value of 0.1 is lower than the IoU value using in our previous work of Mask R-CNN to detect dislocation loops and black dot defects (IoU = 0.3)[2] but comparable to a YOLO model of dislocation loops (IoU = 0.15).[3]

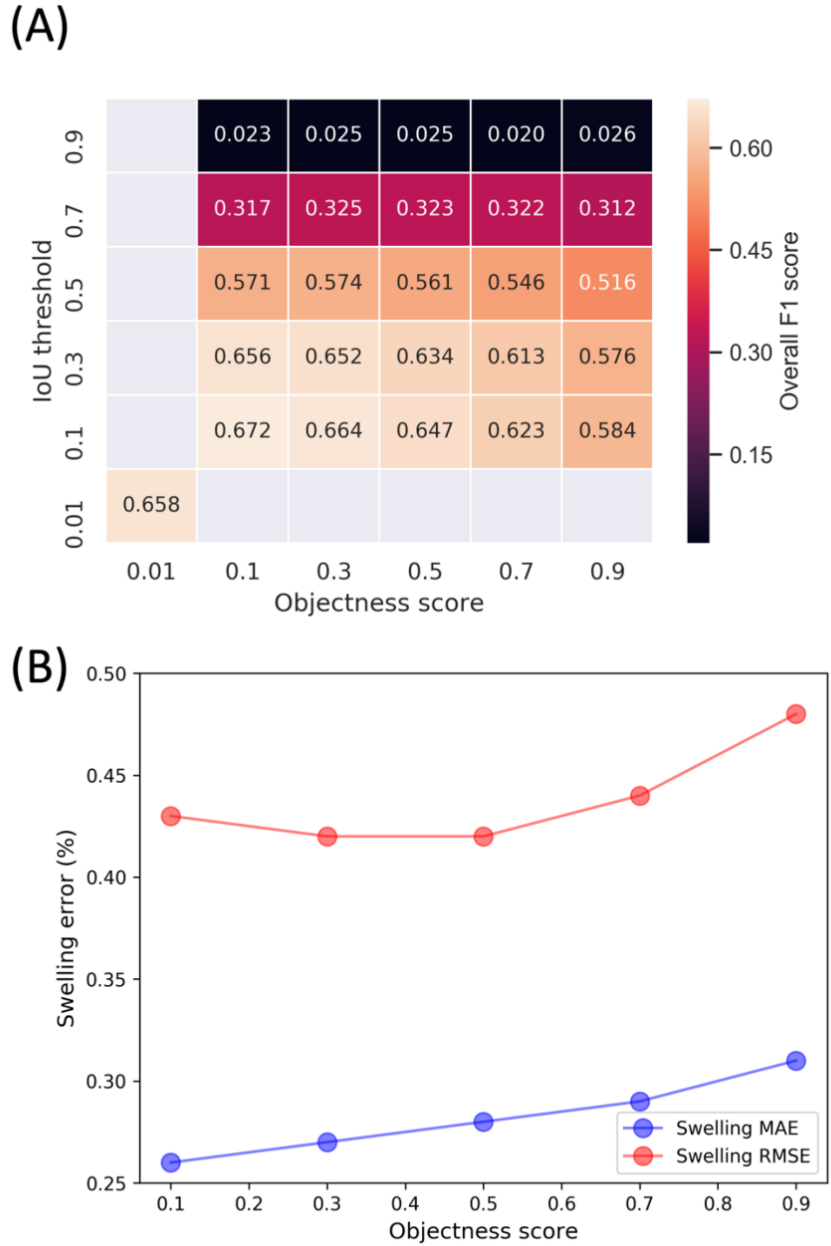

**Figure S10:** (A) Heatmap showing hyperparameter selection to optimize model prediction of material swelling based on the choice of IoU threshold and model objectness score. The heat values correspond to the overall F1 score of the model, where a value of IoU=0.1 and 0.1 objectness score corresponds to the highest overall F1 of 0.672. (B) The mean absolute error (MAE, blue points) and root mean squared error (RMSE, red points) of material swelling as a function of objectness score, with an IoU=0.1. Here, an objectness score of 0.1 results in the lowest swelling MAE of 0.26 percent swelling, and a corresponding RMSE of 0.43 percent swelling. The CNL+NOME initial split was used in these evaluations. Note that lower IoU threshold and model objectness scores of 0.01 resulted in worse performance than the IoU=0.1 and objectness score of 0.1 shown here.

**References**

[1]   C.M. Anderson, J. Klein, H. Rajakumar, C.D. Judge, L.K. Béland, Automated Detection of Helium Bubbles in Irradiated X-750, Ultramicroscopy. 217 (2020) 113068. doi:10.1016/j.ultramic.2020.113068.

[2]   R. Jacobs, M. Shen, Y. Liu, W. Hao, X. Li, R. He, J.R.C. Greaves, D. Wang, Z. Xie, Z. Huang, C. Wang, K.G. Field, D. Morgan, Performance and limitations of deep learning semantic segmentation of multiple defects in transmission electron micrographs, Cell Reports Phys. Sci. (2022) 100876. doi:10.1016/j.xcrp.2022.100876.

[3]   M. Shen, G. Li, D. Wu, Y. Yaguchi, J.C. Haley, K.G. Field, D. Morgan, O. Ridge, O. Ridge, A deep learning based automatic defect analysis framework for In-situ TEM ion irradiations, Comput. Mater. Sci. 197 (2021) 110560. doi:10.1016/j.commatsci.2021.110560.
38